\documentclass[11pt,twoside]{article}
\usepackage{asp2014}

\aspSuppressVolSlug
\resetcounters

\begin{document}

\title{Prospects for unseen planets beyond Neptune}
\author{Renu Malhotra 
\affil{The University of Arizona, Tucson, Arizona, USA; \email{renu@lpl.arizona.edu}}
}

\paperauthor{Sample~Author1}{Author1Email@email.edu}{ORCID_Or_Blank}{Author1 Institution}{Author1 Department}{City}{State/Province}{Postal Code}{Country}

\begin{abstract}
Several anomalous features in the orbital distribution of distant Kuiper Belt objects have recently sparked suggestions of an unseen large planet (or two) orbiting beyond Neptune. I review the theoretical arguments and the prospects for the existence of unseen massive bodies in the distant solar system. 
\end{abstract}

\section{Introduction}

I am very happy to participate in this Symposium to honor Prof. Wing Ip on his seventieth birthday. I met Prof. Ip for the first time only in the summer of 2016, but more than two decades prior one of his papers provided a crucial missing link for my explanation of the peculiar properties of Pluto.

During my graduate studies on resonance dynamics problems in the satellite systems of the giant planets, I learned -- under Stan Dermott's mentorship -- about mean motion resonance capture by means of slow dissipative processes. With my interest in Pluto stimulated by Andrea Milani and Anna Nobili (who were spending a sabbatical year at Cornell, working on their LONGSTOP Project, one of the first long numerical integrations of the outer solar system), it seemed plausible to me that Pluto had been captured into the orbital resonance with Neptune by means of some slow dissipative process that brought the two planets' orbits closer together. We explored the possibility of gas drag on Pluto providing the necessary dissipation and found that although it was fairly easy to get Pluto captured into resonance with Neptune, it was very difficult to excite its eccentricity and inclination, and so we dropped the problem. It did occur to me then that moving Neptune towards Pluto would be a better solution to the puzzle, since it would not damp Pluto's eccentricity, but there was no known mechanism to move Neptune outward to warrant pursuing this idea further.

I was able to revisit this idea a few years later, when I accidentally came across the paper by \cite{Fernandez:1984} in which they reported their computer modeling of the formation of the outer planets, Uranus and Neptune. This was amongst the earliest attempts at numerical simulations of planet formation. Although their computer model was only a rather rough approximation of the physics of planetary accretion, Fernandez and Ip made a very important insight which they described in a heuristic way: as the ice giants grew in mass by accreting planetesimals, Neptune would tend to migrate outward whereas Jupiter would tend to migrate inward as the source of planetesimals was gradually eroded from the inside-out. For almost a decade this paper received rather little attention. But upon thinking about it, I became convinced that it offered a plausible mechanism for moving Neptune outward.

One could now postulate that Pluto had formed in a nearly circular, nearly co-planar orbit in the natal solar system's disk, just like the other planets, but had been swept by Neptune's exterior 3:2 orbital resonance as Neptune migrated outward: the subsequent migration would naturally lead to Pluto's eccentric resonant orbit. Moreover, Pluto's measured orbital eccentricity provided a lower limit on the magnitude of Neptune's outward radial migration (of at least $\sim$5 au). And, by conservation of energy and angular momentum principles, it was also possible to infer the magnitude of the inward migration of Jupiter and of the mass of the planetesimal disk needed to fuel the migration of the giant planets. All of these quantitative estimates did not seem outright implausible, so I wrote up this idea in a short paper \citep{Malhotra:1993}, followed up with another paper \citep{Malhotra:1995} to flesh out some predictions of such planet migration on the dynamical structure of the Kuiper Belt which had just been discovered \citep{Jewitt:1993}. 

I suppose the timing was just right, because the idea of planet migration was then very quickly and widely accepted as a new paradigm for the origin and evolution of the Solar system. It is probably no exaggeration to state that that insight in \cite{Fernandez:1984} has transformed almost all aspects of planetary science in the past 25 years.

In the present contribution, I will discuss some new developments in outer solar system studies: the possibility of unseen planets yet to be discovered.  Let me start by noting that our ancestors many millennia ago recognized five planets (Mercury, Venus, Mars, Jupiter and Saturn).  When William Herschel and Caroline Herschel discovered a new planet in 1781 CE, a planet unknown to the ancients, subsequently named Uranus, the discovery truly opened human imagination to the possibility of more planets. Uranus was discovered serendipitously, in the dawn of the age of telescopes. The next new planet, Neptune, was a planet that was {\it predicted}, motivated by mathematical calculations based on Neptune's theory of gravity which revealed anomalies in the orbital motion of the planet Uranus. There were further predictions of yet another planet beyond, based on perceived anomalies of the motion of Neptune. The search for that predicted planet led to the discovery of Pluto. (A brief, accessible account of this history is given by \cite{OConnor:1996}.) Now, about 87 years later, we recognize that Pluto is one of a number of dwarf planets and many smaller bodies comprising the Kuiper Belt.  But our thirst for more planets has not ceased. However, instead of using anomalous planetary orbital motions to probe for unseen distant planets, we now use the minor planets --the Kuiper Belt objects (KBOs) -- as tracers. We use statistical arguments and we use the geometry of orbits to sense the presence of and even predict the location of unseen planets.  This is a different kind of mathematical analysis for hunting new planets.

\section{Kuiper Belt objects as tracers of unseen planets}

\begin{figure}[!ht]
\centering
\includegraphics[scale=0.6]{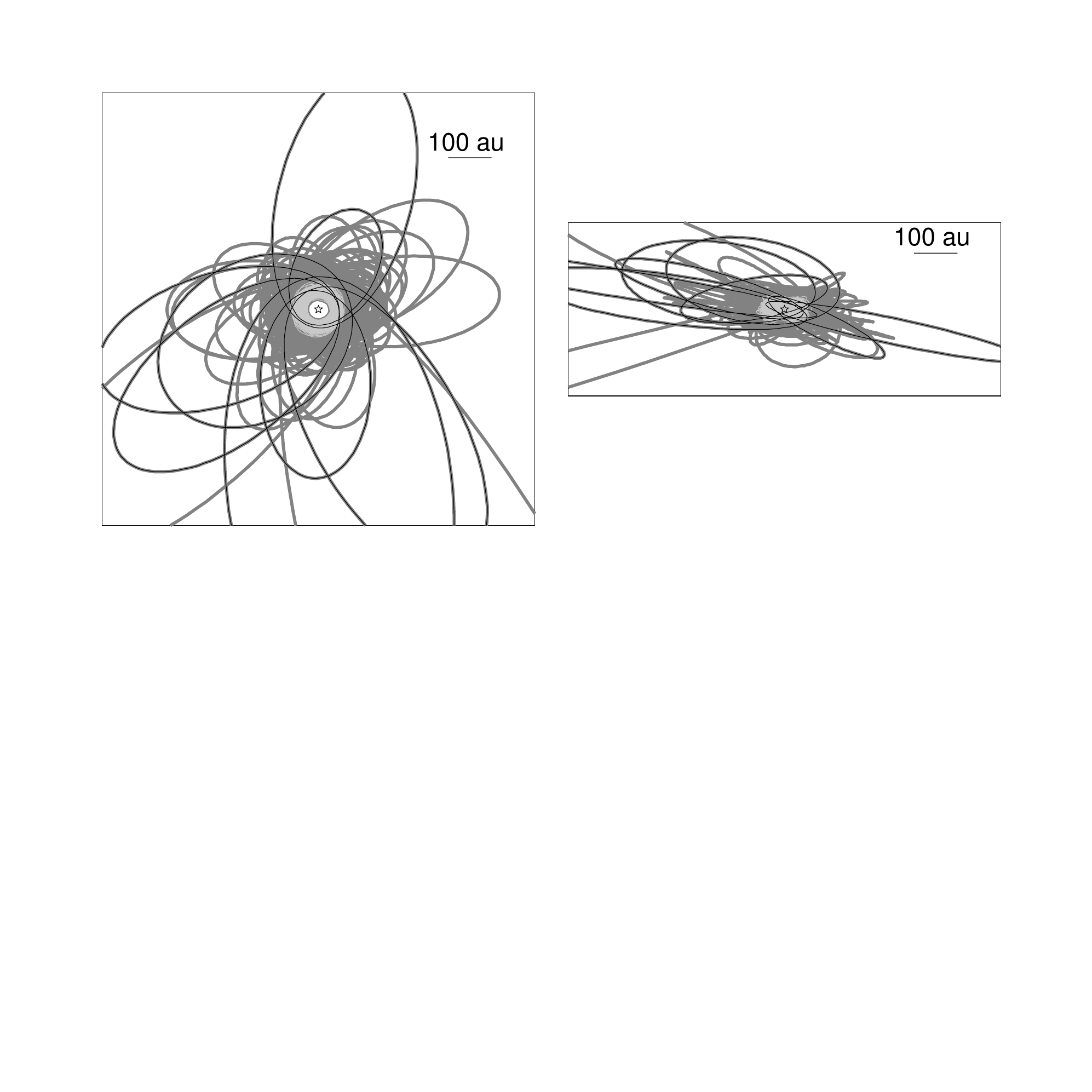}
\vspace{-2.3truein}
\caption{Heliocentric orbits of currently known Kuiper Belt objects (as of February 2017), projected top-down (left) and edge-on (right) to the ecliptic.  Those in light grey are classified as either "resonant" or "classical" objects; there are eight objects traced in black -- so-called "extreme KBOs" -- whose perihelion distance exceeds 40~au and semi-major axis exceeds 150~au.}
\label{f:figxyxz}\end{figure}

In the past two-and-a-half decades, astronomers have discovered more than two thousand minor planets beyond the orbit of Neptune.  This number is comparable to the number of asteroids that were known a century ago, and it is fair to say that our knowledge of the Kuiper Belt is at an early stage, comparable to our knowledge of the asteroid belt as of a century ago.  Figure~\ref{f:figxyxz} shows a top-down and edge-on projection of the relatively well-determined KBO orbits.  We see that the innermost part of the Kuiper Belt consists of moderately eccentric orbits whereas the outer parts, beyond about 50 au, the orbits are highly eccentric as well as highly dispersed in inclination to the ecliptic. The former population, consisting of the resonant objects and the so-called "classical disk" objects, is thought to be the remnants of the in-situ planetesimals that formed locally (at heliocentric distances of 25--47 au), whereas the latter population, consisting of the "scattering" and "scattered" populations, is thought to be survivors of gravitational scattering with Neptune~\citep{Levison:1997}.  The total mass of the Kuiper Belt, including the scattered disk component, is estimated to be only a few times the mass of the Moon~\citep{Gladman:2001,Iorio:2007}.

Given its low total mass and its low number density in the volume of space that it orbits within, the effects of self-gravity of the Kuiper Belt are expected to be negligible, so the KBOs behave as tracer particles, responding to the effects of the gravity of the large planets as well as recording their dynamical history.

\subsection{Mid-plane of the Kuiper Belt}

One of the basic questions we can ask is: what is the mid-plane of the Kuiper Belt and how well does it conform to theoretical expectations?  Within our current knowledge of our planetary system, the known planets are dominant in controlling the dynamics of KBOs, so we expect that the KBOs' mid-plane is close to the nearly common plane of the giant planets, Jupiter--Neptune.  A high accuracy measurement of the Kuiper Belt's mid-plane potentially provides a sensitive probe of unseen planets that might cause deviations of the mid-plane from that enforced by the known planets.  

We can define the theoretically expected mid-plane more precisely as follows.  We can measure the total angular momentum of the known solar system (which is dominated by the orbital angular momentum of the major planets about the solar system's barycenter); the plane perpendicular to the total angular momentum vector is known as the ``invariable plane'' of the solar system.  The orientation of the invariable plane has been measured to better than 1 milli-arcsecond accuracy~\citep{Souami:2012}.  The giant planets orbit the Sun almost, but not exactly, in this plane; the offsets are small but measurable.  Because of these small deviations of the orbital planes from the invariable plane, the mid-plane of the KBOs is also expected to be slightly offset from the invariable plane. 

Examining this question more deeply, we note that while the invariable plane is fixed in time (neglecting external torques), the mutual gravitational effects of the giant planets cause their orbital planes to precess and wobble about the invariable plane in a manner that can be calculated, in the linear approximation, with the Laplace-Lagrange secular theory~\citep{Murray:1999SSD}.  The time-variability of the orbital planes of the giant planets means that the mid-plane of the KBOs can also be expected to be time-variable; the timescale of this variability is $\sim$~megayears. Additionally, the mid-plane of the KBOs can be expected to be not exactly a flat sheet, but to vary with heliocentric distance because the gravitational effects of the planets also change with distance.  All these effects can be calculated with an extension of the same Laplace-Lagrange secular theory~\citep{Murray:1999SSD}.  

The first attempted measurement of the mid-plane of the KBOs was carried out by \cite{Collander-Brown:2003} who focussed their study on the so-called classical KBOs with semi-major axes in the range 40--47~au, and reported that the average orbital angular momentum of their sample (of 141 classical KBOs at the time) was consistent with their mean plane being very close to the invariable plane; these authors did not assess the measurement error, however. Shortly thereafter, \cite{Brown:2004b} reported a measurement of the mean plane of the whole sample of KBOs then known (a sample of 728 objects), finding that the mid-plane was distinctly different than the invariable plane and also distinctly different than the orbital planes of Jupiter and of Neptune, but was close to the plane predicted by the Laplace-Lagrange secular theory for test particles in orbits of semi-major axis $a=44$~au (the median of the observed sample).  Importantly, \cite{Brown:2004b} recognized that the method of identifying the mid-plane by averaging the unit orbit normal vectors is susceptible to serious systematic errors due to observational biases. Instead, they used a more robust method, namely identifying the mean plane as the plane of symmetry of the KBOs' orbital velocity vectors. They also carried out Monte Carlo simulations of models of the inclination distribution of KBOs to estimate the uncertainty of the measured mid-plane.  Two years later, \cite{Elliot:2005} published a detailed study, based on their large campaign to survey for KBOs, and reported a different conclusion, finding that the Kuiper Belt's mean plane is more consistent with the solar system's invariable plane than with the plane predicted by secular theory for $a=44$~au. A few years later, \cite{Chiang:2008} pointed out that the mid-plane of the Kuiper Belt is not a flat sheet, but it should be significantly warped near $a=40.5$~au due to the $\nu_{16}$ nodal secular resonance.  

\begin{figure}[!ht]
\centering
\vspace{-0.2truein}\includegraphics[angle=270,scale=0.48]{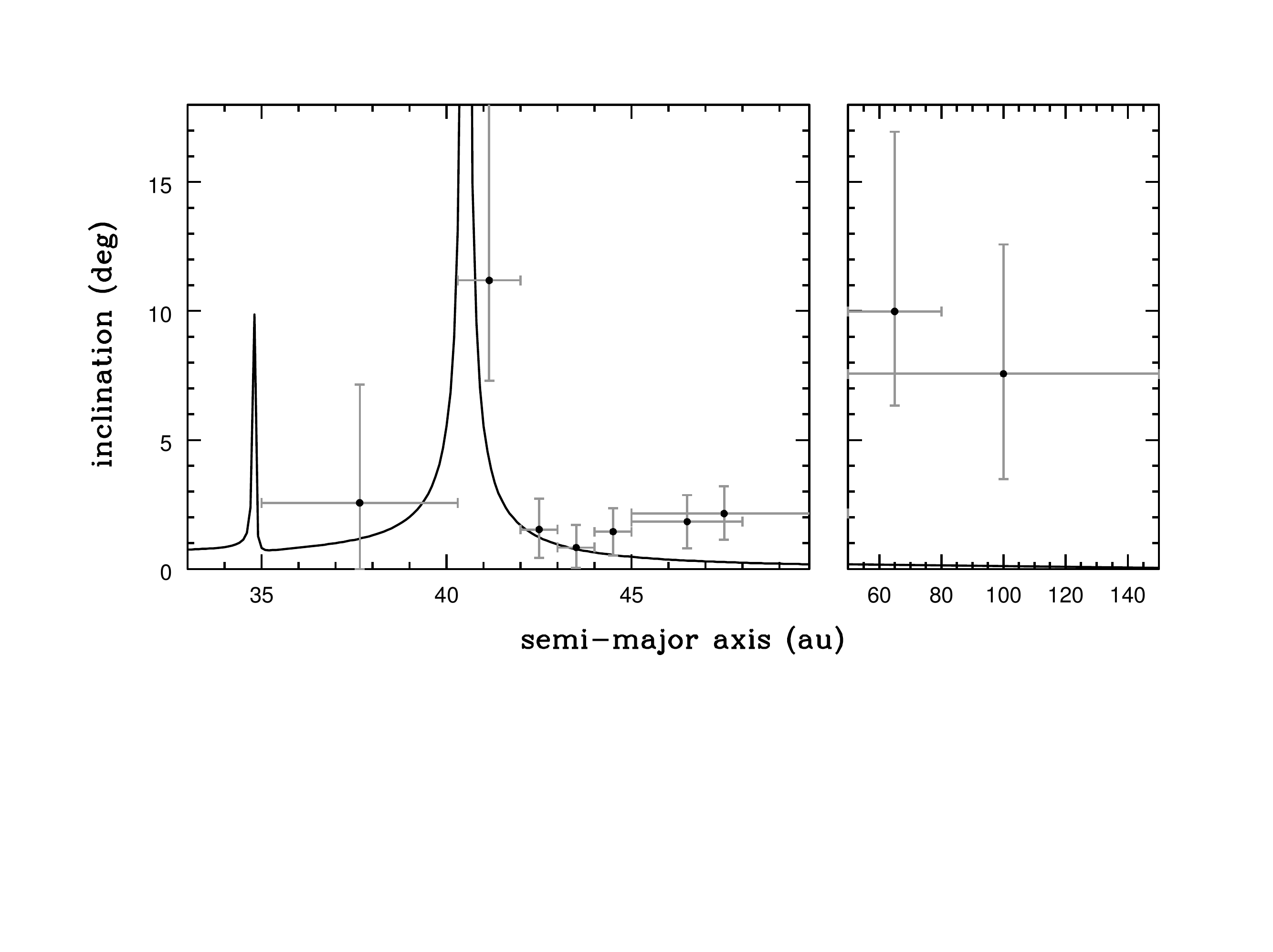}
\vspace{-1.1truein}
\caption{The black curve is the expected mid-plane inclination as a function of semi-major axis, as predicted by linear secular perturbation theory.  The black dots indicate the measured mid-plane inclination of the observed non-resonant KBOs. The vertical grey errorbars indicate the 1--$\sigma$ uncertainty in each semi-major axis bin whose range is indicated by the corresponding grey horizontal errorbar.  The reference plane for the inclinations is the solar system's invariable plane as reported in \cite{Souami:2012}.
(Figure adapted from \cite{Volk:2017}.)}
\label{f:figKBai}\end{figure}

Since the time of the previous measurements of the Kuiper Belt's mid-plane, the observational sample size of KBOs has more than doubled, and the new discoveries encompass a larger range of semi-major axes. 
My colleague, Kathryn Volk, and I have recently examined this data in some detail.  We made a couple of technical improvements to more accurately identify the mid-plane as the plane of symmetry of the orbital velocity vectors of a (biased) sample of KBOs. (One improvement was to identify the unit vector normal to the (unknown) mid-plane by minimizing the sum, $\sum_j |\hat{\bf n}\cdot\hat{\bf v_j}_t|$, of its scalar product with the unit vectors, $\hat{\bf v_j}_t$, directed along the transverse heliocentric orbital velocities of the KBOs.  A second improvement was to carry out all computations with $\sin I$ rather than inclination, $I$, recognizing that the KBOs span a large range of inclinations.)
Our results (reported in \cite{Volk:2017}) are based on the observational sample of those KBOs whose orbits are sufficiently well-determined that their semi-major axis uncertainty is less than 5\% and whose orbits are not resonant with Neptune's; the total number of KBOs in this pared sample is 930. We found that, while the classical KBOs (defined as those in orbits of semi-major axis in the range 35--50~au) have a mid-plane fairly consistent with the predictions of linear secular theory, the mid-plane of the more distant KBOs (those with semi-major axes in the range 50--150~au) is tilted by almost 10 degrees to the expected mid-plane. This result is illustrated in Figure~\ref{f:figKBai}.  The observational sample of distant KBOs is dominated by the Neptune-scattered population known as the "scattered disk objects" which have eccentric orbits with perihelion distances mostly in the range 33--38~au.  Because this sample is not very large and because their orbital planes are highly dispersed (ecliptic inclinations range up to $\sim$~50~degrees), our measurement uncertainty is not small.  However, with Monte Carlo simulations to account for the effects of observational biases with plausible synthetic models of the Kuiper Belt, we demonstrated that the distant KBOs' mid-plane deviates from the expected mid-plane by $>97\%$ statistical significance.

\begin{figure}[!ht]
\centering
\includegraphics[angle=270,scale=0.48]{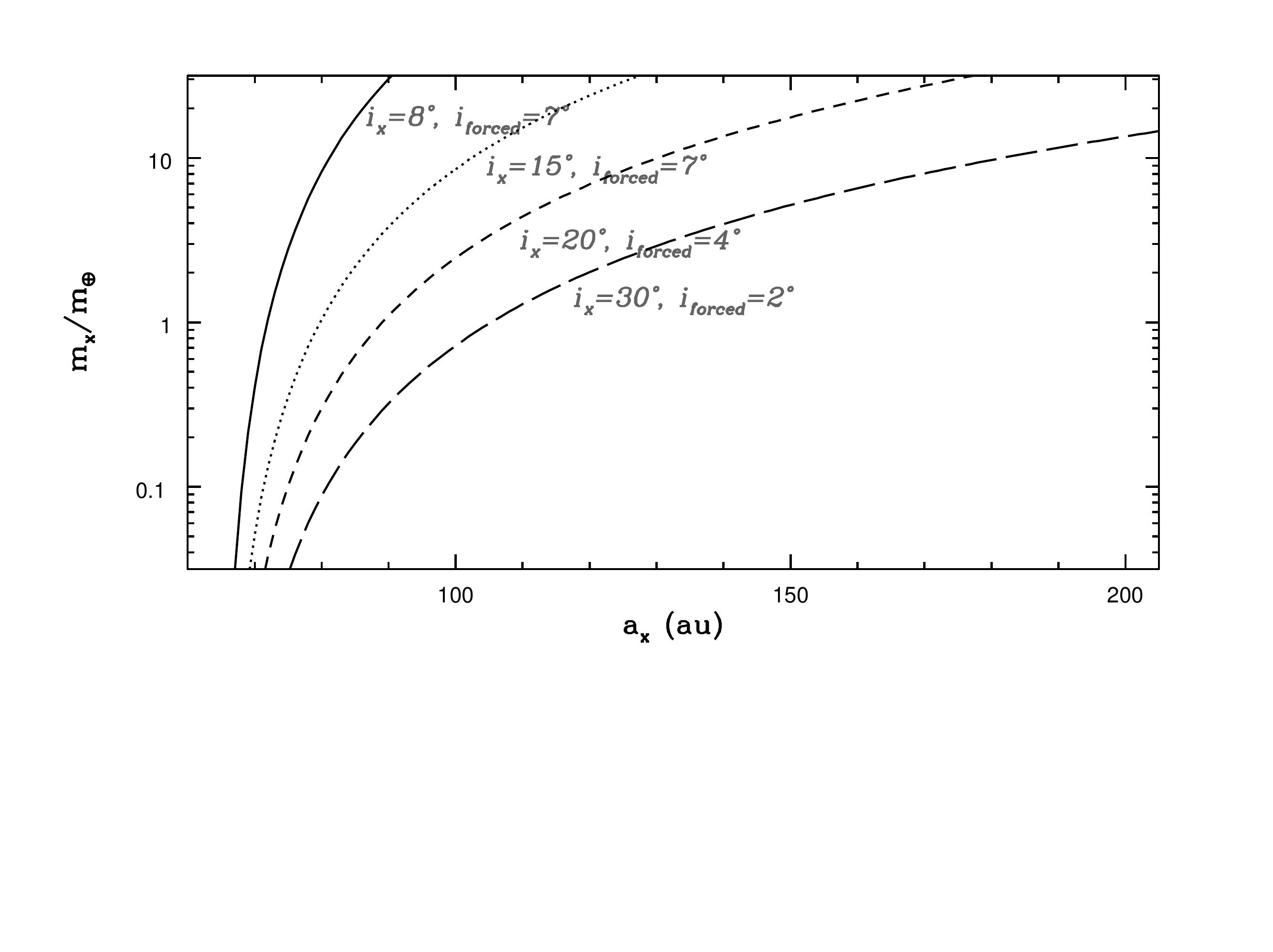}
\vspace{-1.1truein}
\caption{The curves indicate the mass and semimajor axis combinations for a hypothetical planet, with the indicated orbital inclination, $i_x$, which would produce the indicated forced inclination, $i_{\rm forced}$, of test particles at semi-major axis near $65$~au (Figure adapted from \cite{Volk:2017}.)}
\label{f:figXma}\end{figure}

If the mid-plane of the distant KBOs is truly tilted by several degrees, what could account for this tilt?  Considering that this population is affected by gravitational scattering encounters with Neptune over giga-year timescales, one is tempted to consider scattering as a culprit.  However, a few moments' reflection will reveal that the scattering events do not have any preferred plane different than Neptune's own orbital plane.  Furthermore, considering that over timescales of a few megayears, Neptune's orbital plane precesses around the invariable plane, we can expect that the scattered population should have a mid-plane close to the solar system's invariable plane.  Thus, we can reject the hypothesis that Neptune-scattering is the cause of the tilted mid-plane of the distant KBOs.  

Another possible hypothesis is that a close stellar flyby has tilted the mid-plane of the distant Kuiper Belt.  To significantly affect the orbital parameters of KBOs, the stellar flyby would need to be at least as close as about 3--5 times the heliocentric distances of these KBOs.  
Additionally, if a passing star were to perturb the mid-plane of the distant KBOs by $\sim$~10 degrees, subsequent to the stellar encounter, the differential precession of the orbital planes of the population would return their mid-plane to near the invariable plane of the planets on a timescale of 10--30 megayears.  Thus, a stellar encounter at a flyby distance of $\lesssim300$~au would have to have occurred less than $\sim$~30 megayears ago in order to account for the current mid-plane tilt.  While not impossible, a stellar encounter of this nature is of exceedingly low probability~\citep{Garcia-Sanchez:2001}.

The remaining possibility is an unseen planetary mass object on an inclined orbit whose perturbation causes the tilt of the mid-plane of the distant KBOs.  What can we say about such a planet's properties, based only on the measured tilt? Qualitatively, we can expect that the magnitude of the measured tilt depends upon the unseen planet's mass, $m_x$, semi-major axis, $a_x$, and orbital inclination, $i_x$. Quantitatively, this is an inverse problem of solving for the unknown planet's parameters in the secular perturbation theory equations, for a given forced inclination of a test particle.  An approximate analytic solution is derived in Appendix D in \cite{Volk:2017}. In Figure~\ref{f:figXma}, we plot some combinations of $m_x,a_x$ and $i_x$ which yield forced inclinations of the distant KBOs within the 3--$\sigma$ uncertainty range of their measured mid-plane. We can conclude that an unseen Mars-to-Earth mass object orbiting at a heliocentric distance of 60--100~au in an orbit inclined 10--30~degrees to the invariable plane can account for the measured mid-plane of the distant KBOs.  These are approximate bounds. A larger mass at these distances in a lower inclination orbit would account for the mid-plane tilt but is unlikely to have remained undiscovered in previous surveys.  A more distant planetary mass object is not excluded but it would need to be more massive and/or at higher inclination.

\subsection{Other anomalies in the distant Kuiper Belt}

The unexpected tilt of the mid-plane of the distant Kuiper Belt described above is based on the observational sample of KBOs with semi-major axes up to 150 au, because the current observational sample of KBOs of semi-major axis exceeding $\sim$~150 au is very sparse and insufficient for meaningfully identifying their mid-plane.  However, apparent anomalies in the orbital distribution of this small sample have been noted and have recently sparked suggestions of the existence of an unseen "super Earth" planet orbiting at several hundred au.  \cite{Trujillo:2014} reported two features of these so-called "extreme KBOs": they are lacking in objects having perihelion distance in the range 50--75 au, and they exhibit clustering of their arguments of perihelion, $\omega$, near zero.  
These authors suggested, on the basis of numerical simulations, that a massive perturber, a "super Earth" of mass 2--15~$M_\oplus$, in a circular, low inclination orbit of radius $\sim$~250~au could account for the $\omega$ clustering. \cite{Batygin:2016} supported this suggestion by calling attention to the clustering of the longitudes of perihelion as well as the orbital planes of the "extreme KBOs" and arguing, on the basis of a suite of numerical simulations, that an unseen planet of mass $\gtrsim10M_\oplus$ in a distant, eccentric orbit (semi-major axis $\sim$~700~au, eccentricity $\sim$~0.6) would account for the orbital clustering.  With more extensive numerical simulations, \cite{Brown:2016} bounded the unseen planet's parameters to approximately 5--20~$M_\oplus$, semi-major axis 380--980~au, eccentricity 0.56--0.75, inclination 20--30~degrees to the ecliptic.  \cite{Batygin:2016} and \cite{Brown:2016} noted that in their numerical simulations the orbital clustering of the extreme KBOs appeared to be related to the phase-space confinement enforced by mean motion resonances with the unseen planet, but they did not attempt to identify any such mean motion resonances between the observed extreme KBOs and the hypothetical unseen planet.

In \cite{Malhotra:2016}, we noted that the extreme KBOs have orbital period ratios with each other that are intriguingly close to ratios of small integers, and that rational ratios between pairs of KBOs would have dynamical significance only if each KBO were in a mean motion resonance (MMR) with a massive perturber.  Significantly, the closeness to resonant period ratios supports the distant planet hypothesis {\it independent} of the orbital clustering evidence.  However, the rational numbers are dense on the number line, so this observation for a handful of objects has only marginal statistical significance, given the current measurement uncertainties of the orbital periods of the extreme KBOs, as we demonstrated. Nevertheless, we proceeded to examine how resonant relationships with the hypothetical planet might be useful in providing constraints on the planet's properties.  To do this, we needed theoretical understanding of mean motion resonances in the uncharted territory at extremely high eccentricities because textbook treatments have hitherto been confined to low-to-moderate eccentricity orbits, and it is presumed that at high eccentricities the resonance zones dissolve into chaos owing to "resonance overlap" (e.g., \cite{Murray:1999SSD}).  We quickly discovered that, in the simplest planetary model of the planar circular restricted three body problem, the phase space of small integer mean motion resonances has novel features at high eccentricities that allow previously unforeseen geometries of resonant orbits that are not chaotic and are quite stable.   \cite{Wang:2017} describe in detail these novel features for  the 2:1 and the 3:2 exterior MMRs.

\begin{figure}[!ht]
\centering
\includegraphics[angle=270,scale=0.48]{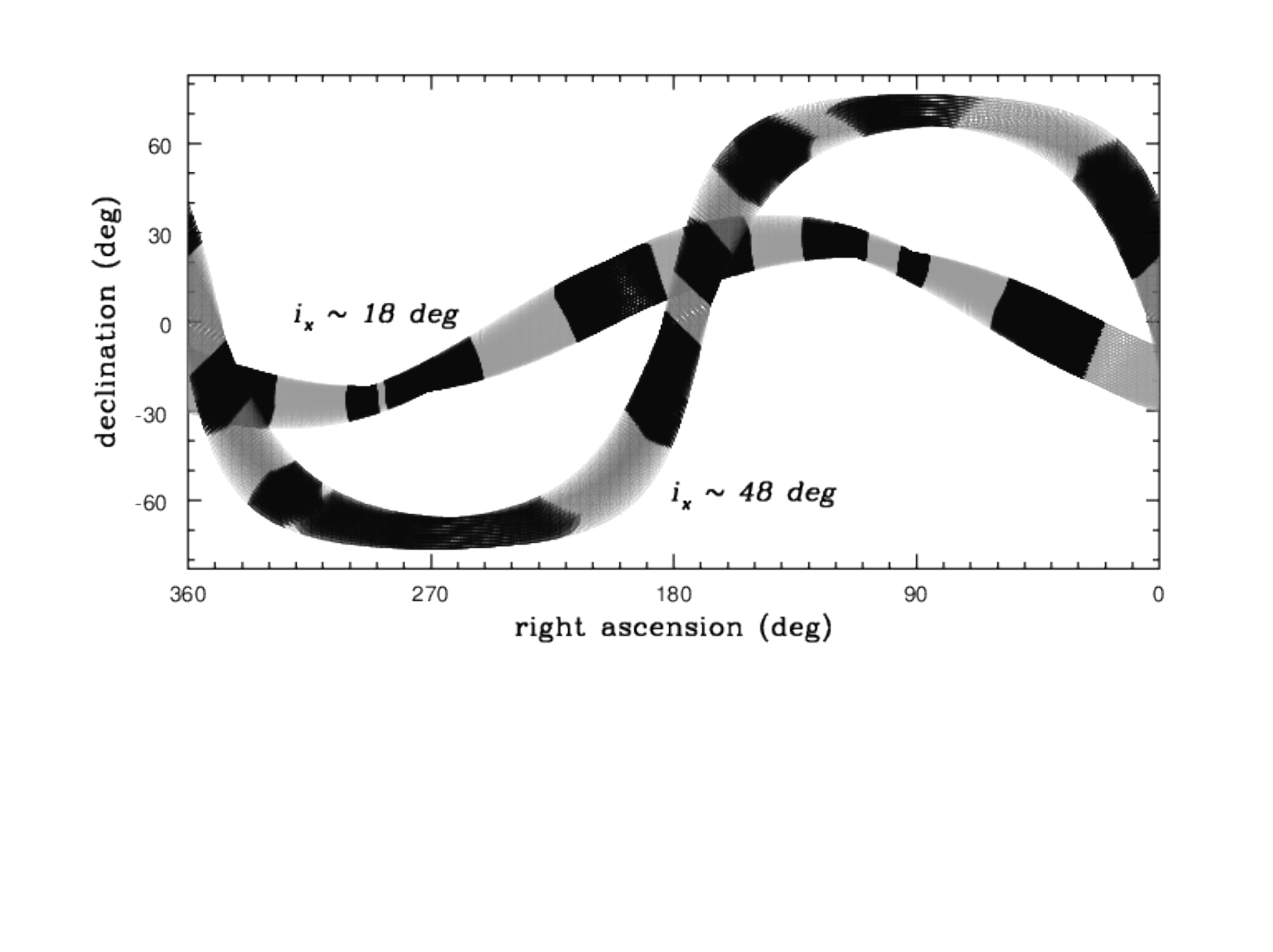}
\vspace{-1truein}
\caption{Possible locations in the sky of the distant hypothetical planet, for low inclination orbits near the mean plane of the extreme KBOs, and for high inclination orbits which place the extreme KBOs near Poincar\'e's periodic orbits of the third kind; the zones in black are excluded by the resonant objects. (Figure adapted from \cite{Malhotra:2016}.)}
\label{f:figsky}\end{figure}

By means of these insights, we hypothesized that a distant planet in an orbit of semi-major axis 664~au could simultaneously hold four of the most extreme KBOs in low integer MMRs.  Instead of viewing the KBOs' orbital period uncertainties as a difficulty for this hypothesis, we reasoned that a sufficiently massive planet would have resonance widths large enough to encompass the period uncertainties; this led us to a lower bound on the planet mass of $\gtrsim10M_\oplus$.  (Interestingly, our independent estimate based on MMR dynamics is not dissimilar to the mass estimates of \cite{Trujillo:2014} and \cite{Brown:2016} based on numerical simulations of orbital clustering.) We also estimated an upper bound on the planet's orbital eccentricity, $\lesssim$~02--0.4, based on the resonant orbits' geometries; however, this is a strict bound only for co-planar orbits, and would be relaxed for non-co-planar orbits.  We also identified a highly inclined orbit of the hypothetical planet, inclined $\sim48^\circ$ to the ecliptic, which would allow each of the four extreme KBOs to have nearly stationary argument of perihelion (relative to the planet's orbit plane); with such an orbit of the hypothetical planet, the four extreme KBOs would be in libration about a periodic orbit of the third kind, a class of periodic orbits of the three dimensional restricted three body problem.  The very long term stability of such librations would likely be reduced in the presence of the perturbations of other planets in the solar system, but the librations would likely persist for secular timescales on the order of tens-to-hundreds of megayears in the distant solar system.  Such dynamical behavior has been noted in the literature on numerical simulations of the Kuiper Belt but has been somewhat erroneously referred to as "Kozai within MMR" (e.g., \cite{Gomes:2005b, Volk:2011, Nesvorny:2015}).  Perhaps most useful for the observational discovery of the unseen planet, we identified the planet's range of current orbital longitude which is excluded by the resonant geometry of each of the four extreme KBOs. Figure~\ref{f:figsky} plots the results of this analysis for the allowed sky location of the unseen planet.

\section{Epilogue}

Currently, the observational census of small bodies in the solar system beyond Neptune is at a stage comparable to what our knowledge was of the asteroid belt about a century ago.  But certain dynamical features are already very evident and have led to a dramatic paradigm shift in our theoretical understanding of solar system history.  The population of KBOs in orbital resonance with Neptune is widely interpreted as evidence of the outward migration of Neptune in the early history of the solar system, as first intimated in the work of \cite{Fernandez:1984} and \cite{Malhotra:1993,Malhotra:1995}.  As observational discoveries have advanced to more distant objects beyond the strong influence of the known planets, the distant KBOs beyond about 50~au are now employed as tracers of possible unseen planetary mass objects beyond Neptune.  The measured large tilt of the mid-plane of scattered disk KBOs (of semi-major axis in the range 50--80~au and perihelion distance 33-38~au) is suggestive of a Mars--to--Earth mass unseen planet orbiting in a moderately inclined plane in that region \citep{Volk:2017}.  The apparent clustering of several orbital parameters of the extreme KBOs (those with semi-major axes exceeding 150~au and perihelion distance exceeding 40~au) is intriguing, as is their apparent proximity to low-integer period ratios.  These anomalies suggest the existence of an even more distant and more massive planet, of mass exceeding ten times Earth's mass and orbiting at heliocentric distance of several hundred astronomical units~\citep{Trujillo:2014,Batygin:2016,Malhotra:2016}.  

If one or both of these two different hypothesized planets do exist, how do they "fit in" with our current understanding of the origin of the solar system?  It is plausible that the relatively closer-in Mars--to--Earth mass planet might "fit-in" as simply the upper end of the continuum of the size distribution of objects already discovered in the Kuiper Belt.  It could have formed in the same zone as Uranus, Neptune and the Kuiper Belt objects and would share their compositional provenance and their dynamical history, including the gravitational scattering and orbital migration history inferred from the dynamical structure of the Kuiper Belt.  However, the very distant "super Earth" planet orbiting at a mean heliocentric distance exceeding 380~au, with an apparent void of similar objects in-between Neptune and this body, is somewhat more challenging to accommodate in current models of solar system formation and evolution, although some brave modelers may have found  ways to do so~\citep{Kenyon:2016}. 

In some contrast with the prediction of Neptune in the nineteenth century, the case for unseen planets now  prominently employs statistical analysis and computer modeling of numerous small bodies as tracers of the gravitational effects of massive bodies.  These methods are notable qualitative changes in the way we do this science.  

Finally, I would be remiss if I did not emphasize that the hints that we presently have for unseen planets in the distant solar system are all based on the statistical properties of a relatively small number of minor planets and dwarf planets in the outer solar system.  The statistics of small numbers, which is presently stimulating our imagination, may also be misleading us.  The deviant tilt of the mid-plane of distant KBOs is based on a sample size of less than 200 objects, 
and its statistical significance does not greatly exceed 3--$\sigma$.  The near-resonant period ratios of some of the extreme KBOs are intriguing but also of marginal statistical significance.  \cite{Shankman:2017} suggest that the orbital clustering of the handful of extreme KBOs (numbering less than 10) is simply observational selection bias, but \cite{Brown:2017} disagrees.  
 Be that as it may, it is certainly a puzzling fact that the orbital planes of Kuiper Belt objects are far more dispersed than any solar system formation models have predicted or can account for.  Scattered distant planets may provide a reasonable explanation.


\acknowledgements  The author is deeply grateful to US taxpayers and the state of Arizona taxpayers for research support over many years, and acknowledges funding from NASA (grant NNX14AG93G) and from NSF (grant AST-1312498), and partial travel support from IANCU--Taiwan and from the Marshall Foundation--Tucson for attending this Symposium.

\medskip
{\footnotesize{This article is based on a presentation made by the author at the symposium, "Serendipities in the Solar System and Beyond", celebrating Wing-Huen Ip's 70th birthday, at the Institute of Astronomy, National Central University, Taiwan, 10--13 July 2017. }}

\end{document}